\documentclass[aps,prb,twocolumn,showpacs,amsmath,amssymb]{revtex4}
\usepackage{txfonts}
\usepackage{graphicx}
\usepackage{cases}
\usepackage{dcolumn}
\usepackage{bm}

\begin{document}

\title{Metal-terminated Graphene Nanoribbons}
\author{Yan Wang}
\author{Chao Cao}
\author{Hai-Ping Cheng}
\affiliation{Department of Physics and Quantum Theory Project, University of Florida, Gainesville, Florida 32611, USA}

\begin{abstract}
We have investigated structure, electronic, and magnetic properties of metal-terminated zigzag graphene
nanoribbons (M-ZGNRs) by first-principles calculations. Two families of metal terminations are studied: 1)
$3d$-transition metals (TM) Fe, Co, and Ni; and 2) noble metals (NM) Cu, Ag, and Au. All systems have
spin-polarized edge states with antiferromagnetic (AFM) ordering between two edges, except Co-ZGNRs and
Ni-ZGNRs which exhibit negligibly small energy differences between AFM and ferromagnetic (FM) states with the
given ribbon width. In the AFM state the TM terminations transform semiconducting ZGNRs into metallic ones,
while the band gap remains in ZGNR with NM terminations. Ferromagnetic states of M-ZGNRs with TM terminations
show a high degree of spin polarization at the Fermi energy. We predict a large magnetoresistance in Fe-ZGNR
junctions with a low, uniform magnetic switching field.
\end{abstract}


\maketitle

\section{\label{sec:1}INTRODUCTION}

Giant magnetoresistance (GMR) \cite{Baibich:1988,Binasch:1989} and tunneling magnetoresistance (TMR) effects \cite{Miyazaki:1995,Moodera:1995,Butler:2001,Parkin:2004,Yuasa:2004}
in spin-valves and magnetic tunnel junctions have been the focus of spintronics studies in the past two
decades. Conventional GMR and TMR devices, based on inorganic multilayered structures whose critical element
is a sandwich of two magnetic layers separated by nonmagnetic metallic or insulating layers, exhibit a large
change in conductance when the orientation of the magnetic layers is changed from parallel (P) to
antiparallel (AP). The magnetic layers are usually made of transition metal ferromagnets, which gives a high
spin polarization at the Fermi energy.

Graphene nanoribbons (GNRs) with armchair (AGNR) or zigzag (ZGNR) edges \cite{Nakada:1996,Fujita:1996,Han:2007,Ozyilmaz:2007,Li:2008}, a novel
organic material system produced by cutting graphene along two high-symmetry crystallographic directions,
have recently attracted attention in spintronics. ZGNR is predicted to be a semiconductor with spin-polarized
electronic states localized on the two edges that are individually ferromagnetically (FM) ordered but
antiferromagnetically (AFM) coupled to each other through the graphene backbone \cite{Son:2006}. Son \emph{et
al.}\ \cite{Son-2:2006} show that ZGNRs become half-metallic when an external transverse electric field is
applied across the edges. Kim \emph{et al.} \cite{Kim:2008} propose a spin-valve based on a ZGNR connected by
two ferromagnetic electrodes using a local magnetic field to switch the magnetization of one of the
electrodes. Very recently a magnetoresistance device based on the transformation from the semiconducting
state to the metallic state of the ZGNR has been proposed by Mu\~{n}oz-Rojas \emph{et al.}
\cite{Munoz-Rojas:2009}; an estimated $200 \, {\rm T}$ magnetic field is required to switch the AFM to the FM
state at room temperature. Although these work demonstrate convincingly that spin-polarization and
magnetoresistance can be realized using ZGNRs, the large magnitude of the electric field required to close
the bandgap, the need of local magnetic control or a huge magnetic field, and small spin correlation length
due to the lack of magnetic anisotropy in ZGNRs\cite {Yazyev:2008} are serious limiting factors for future
applications. Moreover, the predicted magnetic states in ZGNRs have not been experimentally observed so far.

Most of the theoretical studies on GNRs for spintronics applications so far only address ribbons with
hydrogen terminations\cite{Son-2:2006,Kim:2008,Munoz-Rojas:2009}. Recent experiments have shown that
graphitic surface curvature can be used as constraint and guide in which metal clusters aggregate to form
linear islands \cite{Schmidt:2008,Kemper:2009}. The presence of metal atoms adsorbed on the graphene
\cite{Chan:2008,Johll:2009,Krasheninnikov:2009,Cao:2010} or GNR \cite{Rigo:2009} changes its electronic and
magnetic properties and thus is certainly of great interest in physics.

In this paper we show that hybrid metal-terminated ZGNRs (M-ZGNRs) can be excellent candidate for spintronic
applications. We studied various M-ZGNRs with terminations of 3\emph{d} transition metals (TM) Fe, Co, Ni and
noble metals (NM) Cu, Ag, Au using first-principles calculations. Compared with hydrogen terminated ZGNR
(H-ZGNR), the TM and NM terminations lead to drastically different and unexpected effects in the structural,
electronic and magnetic properties of the ribbon. A very large spin polarization is predicted for M-ZGNR with
TM terminations in FM state. We propose a new type of junction with a low, uniform magnetic switching field
that utilizes the electronic properties of M-ZGNRs, for which a magnetoresistance value as high as 200\% can
be obtained.

\section{\label{sec:2}computational methods}

The electronic structure calculations are performed using density functional theory (DFT) implemented in the
PWSCF code \cite{Giannozzi:2009}. Each M-ZGNR is simulated within a supercell geometry containing two
pristine ZGNR unit-cells with 40 C atoms and 4 metal atoms at two edges. Ultrasoft pseudo-potentials with
kinetic energy cutoff of 500 eV are employed in all simulations. For the exchange and correlation functional
we used the Perdew-Burke-Ernzerh generalized gradient approximation (GGA) \cite{Perdew:1996}. A vacuum layer
of 15 {\AA} is used between two edges and of 15 {\AA} between two graphene planes to prevent interaction
between adjacent images. Brillouin-Zone sampling uses a grid of 12 Monkhorst-Pack \cite{Monkhorst:1976}
\emph{k}-points along the periodic direction of the ribbon. The optimization of atomic positions proceeds
until the force on each atom is less than 0.01 eV/{\AA}.

The transport calculations of the proposed Fe-ZNGR magnetic junction are performed by using the TRANSIESTA
code \cite{Soler:2002}, which is based on nonequilibrium Green's function formalism and DFT as implemented in
the SIESTA package \cite{Soler:2002}. In order to reproduce the correct band structure of the lead, it is
necessary to use double-$\zeta$ basis with polarization for Fe atoms, while single-$\zeta$ basis with
polarization is sufficient for C atoms. The equivalent plane-wave cutoff for the real-space grid of 150 Ry is
used in throughout the calculations. Current-voltage characteristics within small bias are calculated from
the transmission coefficient $T(E)$ using $I=\int T(E) [f(E-\mu_L)-f(E-\mu_R)] dE$. The magnetoresistance can
therefore be calculated via ${\rm MR}=(I_{\rm P}-I_{\rm AP})/I_{\rm AP}$.

\section{\label{sec:3}Results and Discussion}

\subsection{\label{sec:31}Structure, electronic, and magnetic properties}

\begin{figure}
{\includegraphics[width=8.6cm]{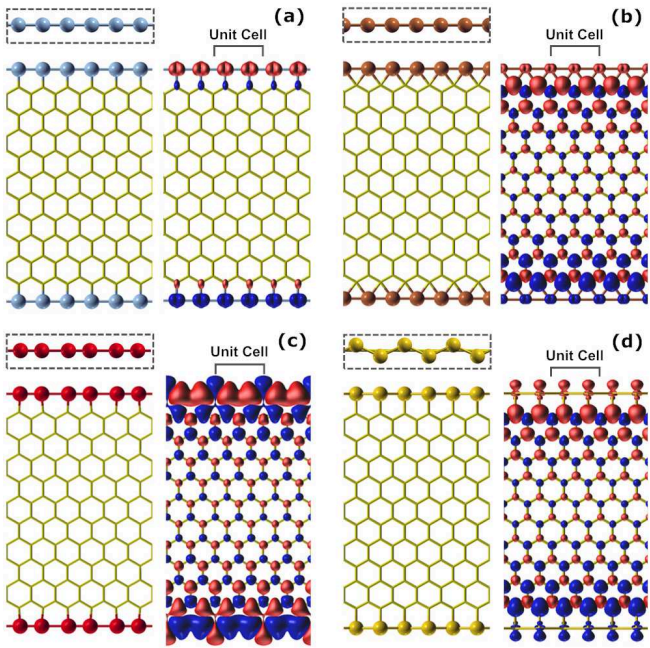}} \caption{\label{fig:geometry}(Color online). Four types
of metal-terminated ZGNR: (a) Ni-ZGNR as linear A type (LA); (b) Cu-ZGNR as linear B type (LB); (c) Fe-ZGNR
as dimerized linear type (DL); (d) Au-ZGNR as zigzag type (ZZ). The atomic configuration appears on the left,
and on the right the corresponding spin density ($\rho=\rho_\alpha-\rho_\beta$) isosurfaces of the M-ZGNRs.
The red (grey) and the blue (dark) isosurfaces denote $\pm 0.007 \, |e|/{\AA}^3$, respectively. The dashed
box shows the edge view.}
\end{figure}

\begin{table*}
\caption{\label{tab:table} Structural, energetic and electronic properties of M-ZGNR. Columns show: metal
termination, structures, metal-metal ($d_{\rm M-\rm M}$) and metal-carbon ($d_{\rm M-\rm C}$) bond length (in
{\AA}), the binding energy $E_b$ and the formation energy of the metal chain $E^{\rm chain}_{\rm formation}$
(in eV per metal atom), energy difference $\Delta E=E_{\rm FM}-E_{\rm AFM}$ between the ferromagnetic and the
antiferromagnetic state (in meV/unit-cell), charge transfer of the metal atom (in $|e|$) from Bader analysis,
total magnetic moment $m_{\rm tot}$ of the ribbon in the FM state (in Bohr magneton $\mu_{\rm B}$ per edge
termination), local magnetic moment of the metal atom $\mu_{\rm M}$ and its nearest-neighbor C atom $\mu_{\rm
C}$ (in $\mu_{\rm B}$), numbers of conduction bands crossing the Fermi level ($n_{\rm AFM}$: the AFM state;
$n_{\rm FM}^{\alpha}$ and $n_{FM}^{\beta}$: majority and minority-spin in the FM state), and spin
polarization $P_{\rm FM}$ of the ribbon at the Fermi energy in the FM state. Results for H-terminated ZGNR
are also listed for comparison.}
\begin{ruledtabular}
\begin{tabular}{cccccccccccccc}
& structures & $d_{\rm M-\rm M}$ & $d_{\rm M-\rm C}$ & $E_b$ & $E^{\rm chain}_{\rm formation}$ & $\Delta E$ &
Charge transfer & $m_{\rm
tot}$ & $\mu_{\rm M}$ & $\mu_{\rm C}$ & $n_{\rm AFM}$ & $n_{FM}^{\alpha}$($n_{FM}^{\beta}$) & $P_{\rm FM}$ \\
\hline Fe-ZGNR& DL & 2.19(2.74)\footnotemark[1] & 1.87 & 4.13 & 1.86 & 6.8 & $-0.31$ & 2.19 & 2.47 & $-0.27$
& 6 & 3(4) & 82 \% \\
Co-ZGNR& LA & 2.46 & 1.82 & 4.26 & 1.79 & 1.3 & $-0.22$ & 1.17 & 1.33 & $-0.11$ & 10 & 3(7) & 78 \% \\
Ni-ZGNR& LA & 2.46 & 1.80 & 4.44 & 1.74 & $<$0.1 & $-0.19$ & 0.18 & 0.23 & $-0.01$ & 8 & 3(5) & 55 \% \\
Cu-ZGNR& LB & 2.46 & 2.07 & 3.42 & 1.37 & 10.1 & $-0.41$ & 0.26 & 0.01 & 0.29 & 0 & 1(1) & $\approx$ 0 \\
Ag-ZGNR& ZZ & 2.74 & 2.15 & 2.38 & 1.20 & 7.8 & $-0.19$ & 0.28 & 0.01 & 0.26 & 0 & 1(1) & $\approx$ 0 \\
Au-ZGNR& ZZ & 2.71 & 2.07 & 3.00 & 1.65 & 11.5 & $-0.08$ & 0.26 & 0.02 & 0.26 & 0 & 1(1) & $\approx$ 0 \\
H-ZGNR& LA & 2.46 & 1.09 & 4.89 & 0.14 & 15.1 & \phantom{-} 0.01 & 0.26 & 0.01 & $-0.30$ & 0 & 1(1) & $\approx$ 0 \\
\end{tabular}
\end{ruledtabular}
\footnotetext[1]{Fe dimerization.}
\end{table*}

To determine the lowest-energy structures of the M-ZGNR, first we carried out structural relaxations for four
possible termination configurations: the linear A type (LA), the linear B type (LB), the dimerized linear
type (DL), and the zigzag type (ZZ), as shown in Fig.\ \ref{fig:geometry}. Both edges of the ribbon have the
same configuration for all systems. The calculated ground-state configurations as well as energetic and
electronic properties are summarized in Table \ref{tab:table}. We find that for all cases the metal atoms
bond strongly with edge carbon atoms, though the bonding configurations are quite different from one another,
as shown in Fig.\ \ref{fig:geometry}. For M-ZGNRs with Co and Ni terminations, the relaxed structures are LA
type. For Cu-ZGNR the most favorable structure is LB type, which is 0.98 eV/unit-cell more stable than the LA
type structure, in good agreement with the result of Wu \emph{et al.} \cite{Wu:2010}. The Fe atoms dimerize
at the edge of Fe-ZGNR due to the Peierls distortion \cite{Peierls:1955}. The most favorable structures for Ag-ZGNR
and Au-ZGNR are ZZ type, with a metal-metal distance larger than the length of pristine ZGNR unit-cell. The
binding energies $E_b$ of the metal atom in M-ZGNRs, defined as $E_b=(E_{\rm M-ZGNR}-E_{\rm ZGNR})/4-E^{\rm
atom}_{\rm M}$, and the formation energy of the metal chain, $E^{\rm chain}_{\rm formation}=E^{\rm
chain}_{\rm M}/2-E^{\rm atom}_{\rm M}$, are shown in Table. \ref{tab:table}. The large differences between
$E_b$ and $E^{\rm chain}_{\rm formation}$ indicate a large direct binding between the metal and carbon atoms
at the ribbon edge. Bader analysis based on the real-space-charge density \cite{Henkelman:2006} for the M-ZGNRs shows charge transfer from metal
atom to C atom for all cases. The amounts of charge transfer for Au-ZGNR and Ag-ZGNR are 0.08 and 0.19
$|e|$/metal-atom, respectively. Such a difference can be important for nanocatalysis.

For each M-ZGNR considered, the ground state has FM spin ordering of the metal atoms at each edge, while AFM
order between the two edges is favored over FM, similar to H-ZGNRs and ZGNRs without H-passivation
\cite{Son:2006}. AFM coupling of the two edges is energetically more stable due to the indirect RKKY exchange
interaction as a result of the bipartite lattice \cite{Saremi:2007}, since the metal terminations always
stand at the opposite sublattices at the two edges. For Fe-ZGRNs, the energy difference $\Delta E$ between
the FM and AFM states increases with decreasing ribbon width \cite{Ong:2010} due to increased interaction
between enhanced spins in inner C sites inside the ribbon. Without the ribbon host the two Fe chains can only
show direct FM exchange coupling through vacuum, and the calculated $\Delta E$ of the two Fe chains becomes
less than 0.1 meV/{\AA} when the distance is larger than 8 {\AA}. Nonetheless, as shown in Table.
\ref{tab:table}, the $\Delta E$ for all M-ZGNRs are less than that of H-ZGNR, indicating a relatively small
effective magnetic coupling between two edges, especially for TM terminations.

In general, the total magnetic moment of the system comes mostly from the metal atoms and their
nearest-neighbor C atoms at the M-ZGNR edges. For TM-ZGNRs, the edge C atom presents magnetization
antiparallel to the nearby TM atom. For Co-ZGNR and Ni-ZGNR in particular (see Fig.\ \ref{fig:geometry}) the
magnetic moments of the ribbon become more localized on the metal atoms at the edges compared to the inner C
atoms in the ribbon, which leads to a negligibly small exchange energy shown in Table. \ref{tab:table}. For
NM-ZGNRs, the metal atom shows magnetization parallel to that of the neighboring C atom at the edge, of order
0.01 $\mu_{\rm B}$/metal-atom, mainly due to the proximity effect. The large difference between magnetic
moments in TM and NM terminations reflects their different physical origins. In TM-ZGNRs, the relatively
large moments of TM atoms come from unfilled $d$-shells and the TM-C covalent bond reduces the number of
unpaired $d$-electrons, giving a reduced moment compared to the freestanding TM atom or monoatomic chain. For
Cu, Ag and Au terminations with filled $d$-shells, the $s$ valence electron saturates the dangling $\sigma$
bonds similar to the hydrogen terminations in H-ZGNRs. The proximity-induced local magnetic moments for Cu,
Ag, Au atoms are negligible, but the total edge moments of about 0.3 $\mu_{\rm B}$/metal-atom are very close
to that of H-ZGNR. Although Mermin-Wagner theorem \cite{Mermin:2006} excludes long-range order in $1D$ spin
lattice model at finite temperature which would limit the dimension of devices based on H-ZGNRs to several
nanometers at room temperature \cite{Yazyev:2008}, improvements can be expected in systems of M-ZGNRs by
increasing the magnetic anisotropy as well as the spin correlation length.

\begin{figure}
{\includegraphics[width=8.6cm]{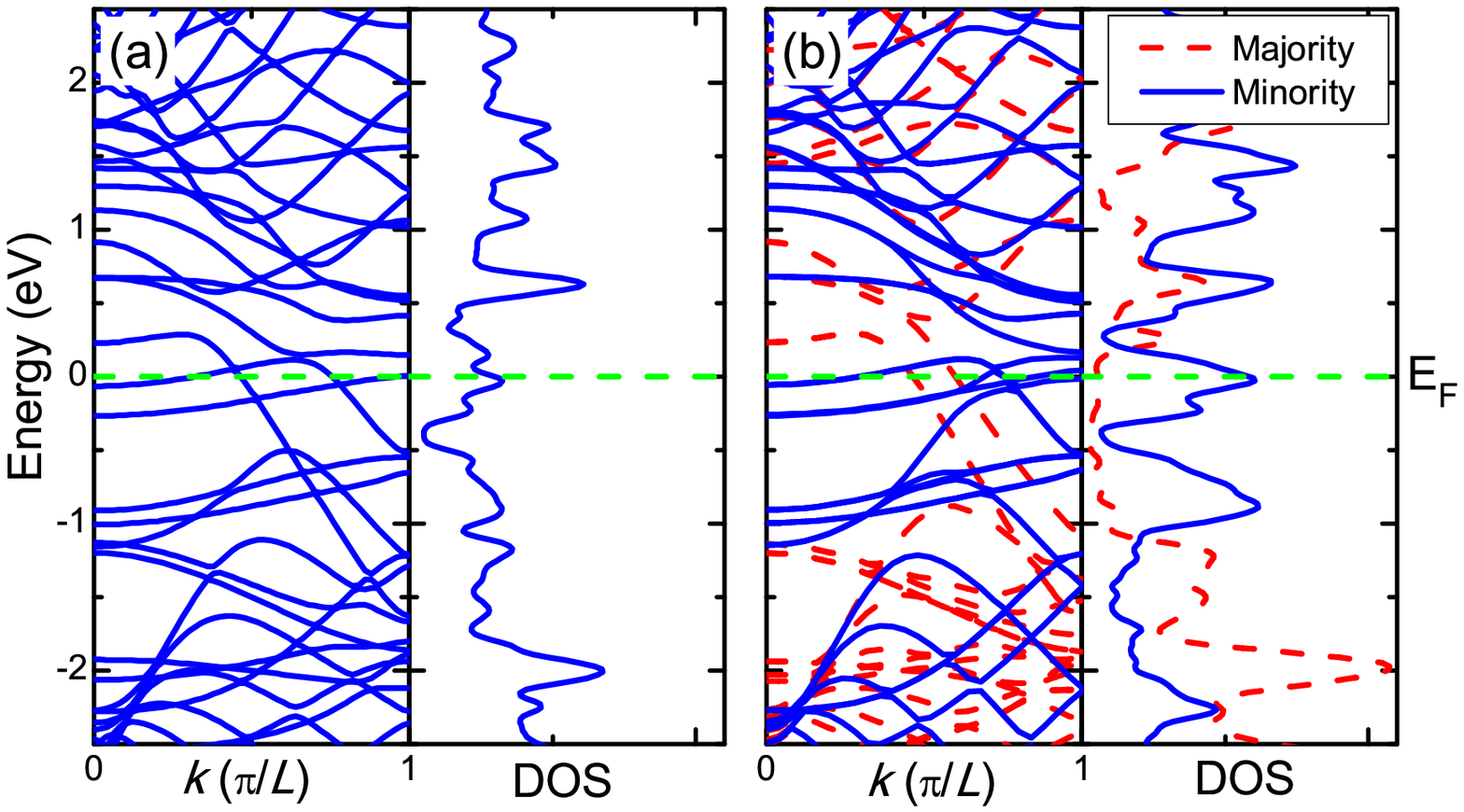}}
{\includegraphics[width=8.6cm]{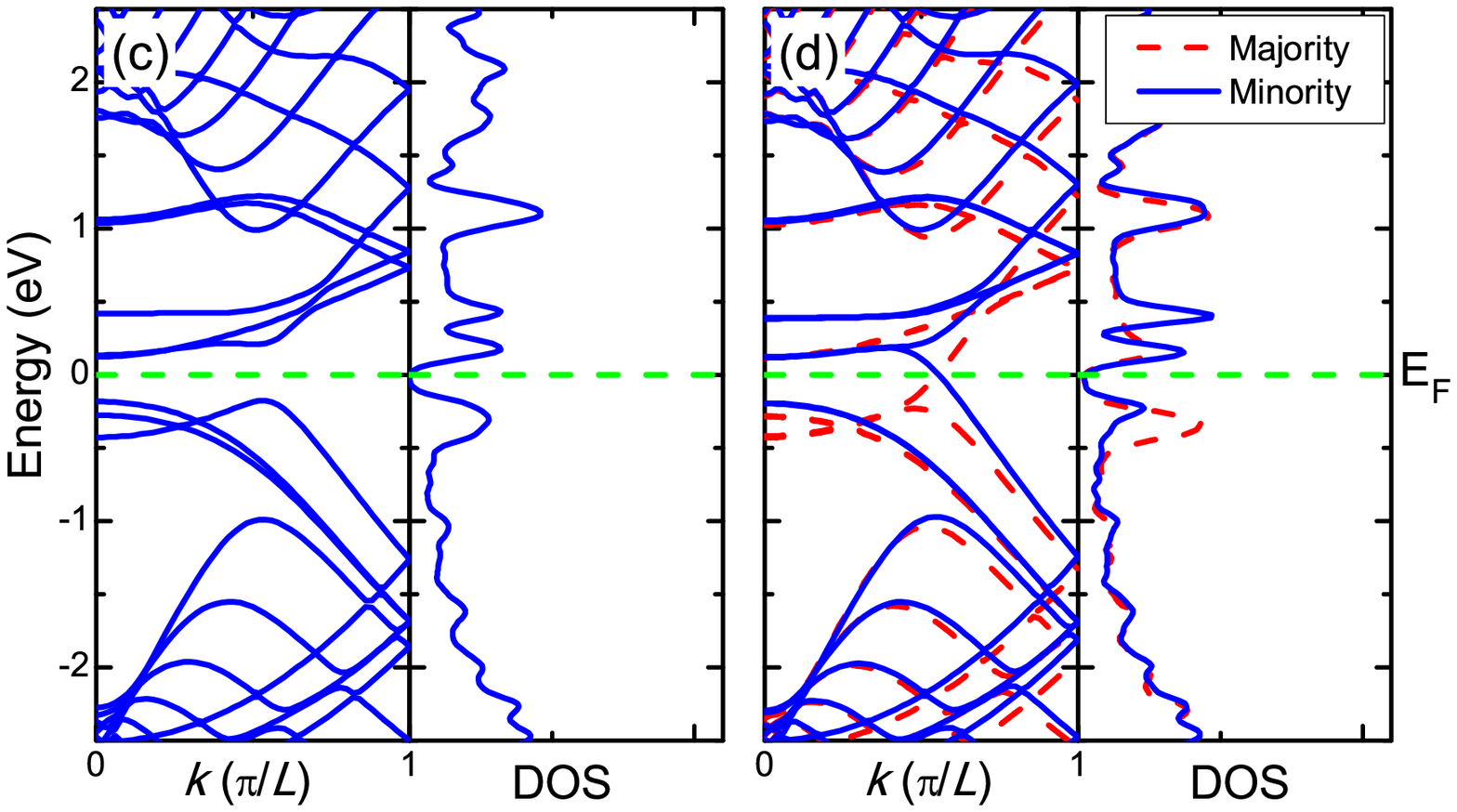}} \caption{\label{fig:band}(Color online). Energy band
structures (left) and density of states (DOS, right, in states/eV) for the M-ZGNRs. (a): Fe-ZGNR in AFM
state; (b): Fe-ZGNR in the FM state; (c) Au-ZGNR in the AFM state; (b): Au-ZGNR in the FM state. The dashed
(red) and solid (blue) lines show the majority and minority spin, respectively. In the AFM state, majority
and minority bands (DOS) coincide.}
\end{figure}

The ground-state TM-ZGNRs are all metallic and the NM-ZGNRs are all semiconductor. We show the band
structures and density of states (DOS) of Fe-ZGNR and Au-ZGNR in Fig.\ \ref{fig:band}. For Fe-ZGNR, both AFM
and FM states are metallic. In the FM state, more flat bands of minority spin cross the Fermi level compared
to those of the majority spin, resulting in a high spin polarization at the Fermi level, as can be seen in
Fig.\ \ref{fig:band} (b). A detailed analysis of the projected DOS indicates that the states near the Fermi
level are mainly from the Fe $d$ orbital. Spin polarization $P_{\rm FM}$ can be calculated as $P_{\rm
FM}=|N_\alpha-N_\beta|/(N_\alpha+N_\beta)$, where $N_\alpha$ ($N_\beta$) represents the DOS of majority
(minority) spins at the Fermi energy. The $P_{\rm FM}$ of the Fe-ZGNR at $E_{\rm F}$ is about 82\% which is
much higher than that of bulk Fe \cite{Tedrow:1973,Soulen:1998}. This implies that in the ballistic transport regime, the
conductance of electrons with one type of spin is larger than the opposite spin. Similar to Fe-ZGNRs, the FM
states of Co and Ni terminated ribbons also show high spin polarization, 78\% and 55\% respectively.

However, all studied NM-ZGNRs, have a bandgap comparable to that of the hydrogenated ribbon (0.32 eV) in the AFM
ground-state. The band structure and density of states of Au-ZGNR as a typical case for NM-ZGNRs are
displayed in Fig.\ \ref{fig:band} (c) and (d). In the FM metallic state the Au-ZGNR shows a large spin
polarization only at the energies slightly away from the $E_{\rm F}$. We also note that the band structure of
NM-ZGNR is very sensitive to the edge structure. The metastable LA type of Au-ZGNR is metallic for both FM
and AFM states.

\subsection{\label{sec:32}Metal-terminated graphene nanoribbon junction}

\begin{figure}
{\includegraphics[width=8cm]{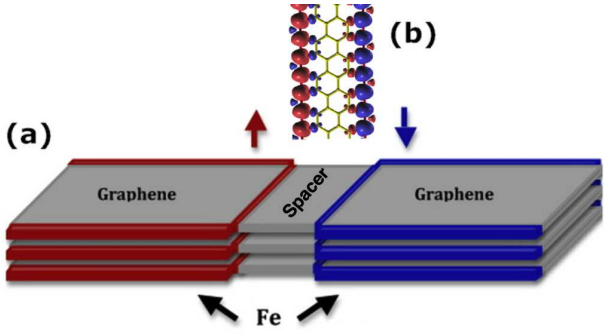}} \caption{\label{fig:junction}(Color online). (a) Schematic
figure of a multilayer Fe-ZGNR/spacer/Fe-ZGNR junction prototype with antiparallel ground-state
configuration. (b) Calculated ground-state (AFM) spin density isosurfaces of the Fe-AGNR with a width of 7.2
{\AA}.}
\end{figure}

Our calculations of truncated Fe-ZGNRs provide an interesting result: if the transverse armchair edge is
terminated by Fe, the strong magnetic coupling will align all edge spins in the same direction. Such highly
spin polarized FM states can be useful in spintronics; in particular, a junction consisting of two TM-ZGNR
leads separated by a spacer can operate as a magnetoresistance device. Fig.\ \ref{fig:junction} (a) is a
prototype multilayer junction stacking three layers of Fe-ZGNRs. The center spacer can be a nonmagnetic metal
such as Cr or Cu, since the behavior of the AFM exchange coupling between magnetic layers through a
nonmagnetic spacer is well-known \cite{Grunberg:2008} and a GMR junction can thus be made; one can also use a
small piece of non-conducting AGNR connected to the two leads via Fe chains, as the spacer which will provide
a magnetoresistance junction. The key issue is the AP configuration being the ground state through the
spacer: the junction can be switched to the P configuration under a uniform external magnetic field applied
to both leads, without the need for a local magnetic field. A conductance difference between P and AP is in
general expected. Our calculations show that the Fe-terminated two armchair edges favor either AFM or FM
configurations depending on the width of the spacer. In Fig.\ \ref{fig:junction} (b) we show an Fe-AGNR with
a width of 7.2 {\AA} which has an AFM ground-state that is 1.1 meV/{\AA} lower than the FM state. AFM
coupling of the two armchair edge states of the spacer can lead to an AP ground-state magnetic coupling
between the two Fe-ZGNR leads. This energy difference can be tuned by increasing the thickness of the lateral
Fe chain and by using multiple layers as shown in Fig.\ \ref{fig:junction} (a). We also note that such a junction can be made by two Fe-AZGNR leads separated by a Fe-ZGNR spacer with similar function \cite{AGNR-junction}.

\begin{figure}
{\includegraphics[width=8.6cm]{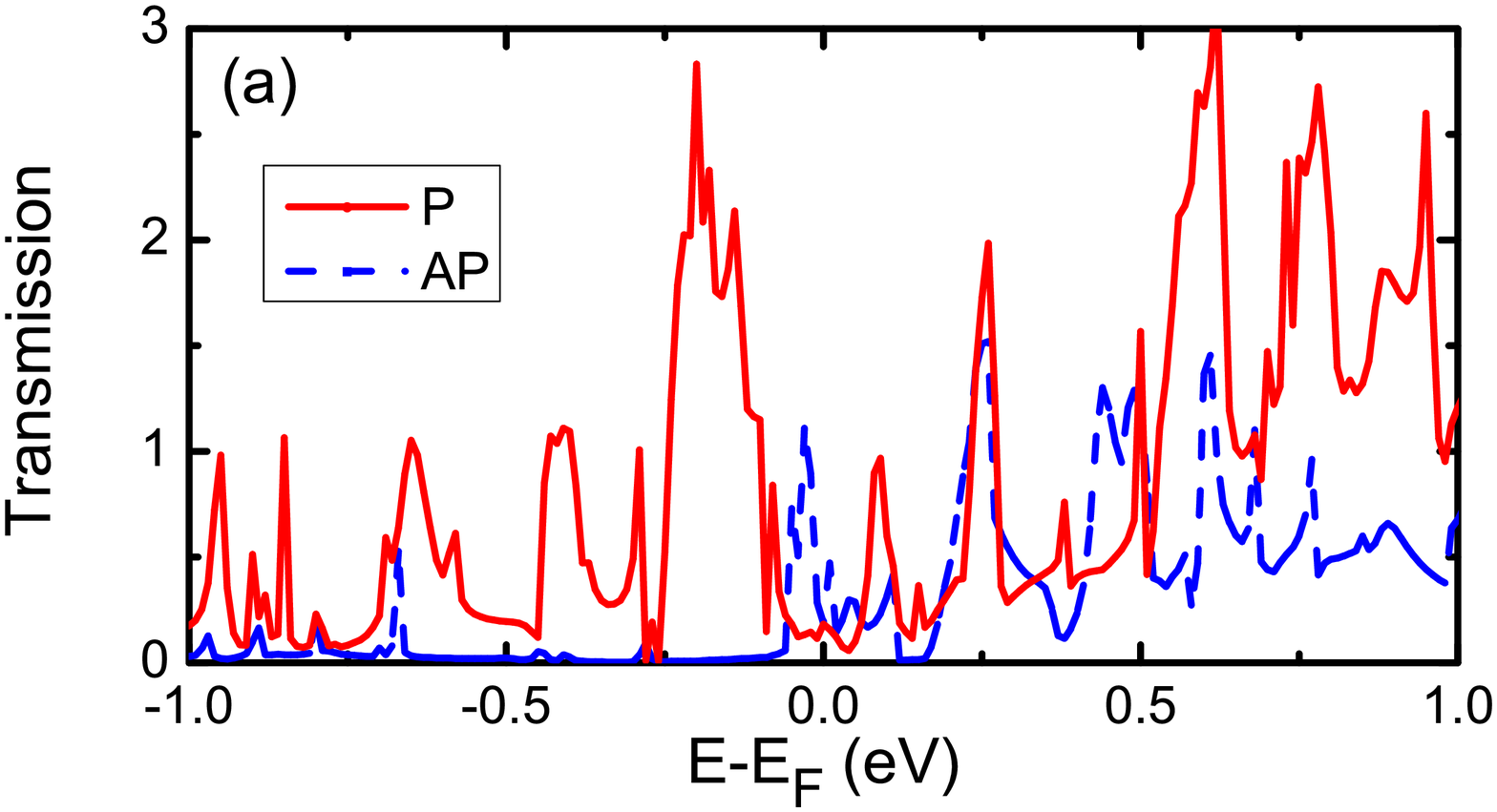}} {\includegraphics[width=8.6cm]{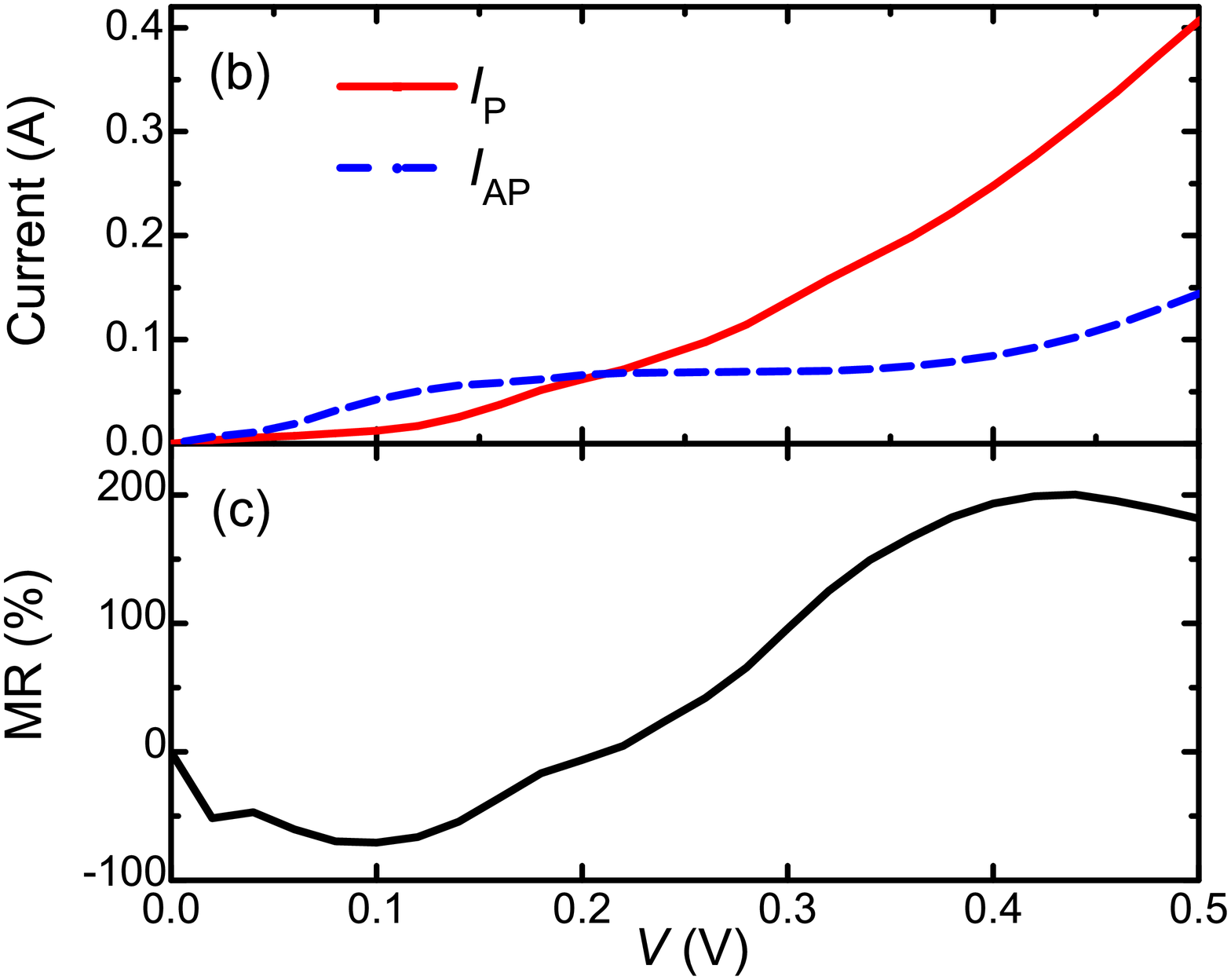}}
\caption{\label{fig:transport}(Color online). Transport properties for a model magnetic tunnel junction with
two semi-infinite Fe-ZGNR leads separated by a 3.1 {\AA} vacuum gap. (a) Energy-dependent transmission
coefficients and (b) current-voltage characteristics within small bias region. The solid (red) and dashed
(blue) lines represent the results for parallel (P) and antiparallel (AP) magnetic configuration of the two
leads, respectively. (c) Calculated magnetoresistance (MR) as a function of bias voltage.}
\end{figure}

To illustrate the possible magnetoresistance, we have performed quantum transport calculations for a model
magnetic tunnel junction based on two semi-infinite Fe-ZGNR leads. A 3.1 {\AA} gap is set between two leads
to represent the Fe-AGNR spacer. The spin-dependent current-voltage characteristics for small bias (linear
response region) are calculated from the energy-dependent transmission coefficients in the ballistic
tunneling regime. Large transmission differences are found near the Fermi energy, as shown in Fig.\
\ref{fig:transport} (a). By comparing the current-voltage characteristics between P and AP configurations in
Fig.\ \ref{fig:transport} (b), we find a magnetoresistance up to 200\% within the small bias region, which is
comparable to the TMR ratio observed in conventional Fe/MgO/Fe sandwiched junctions
\cite{Parkin:2004,Yuasa:2004}. Negative magnetoresistance is also found in the region from 0 to 0.2 eV, as
shown in Fig.\ \ref{fig:transport} (c), primary due to the existence of several large peaks in the
transmission of AP configuration which are very close to the Fermi energy.

Finally we discuss the operational conditions of our proposed magnetic junction of Fe-ZGNR/Fe-AGNR/Fe-ZGNR.
For a lead of infinite size any external field would switch the magnetization of the ZGNR lead. For a
junction with finite-size leads, the critical switching magnetic field is defined as $B=\Delta E/g\mu_{\rm
B}M_{\rm tot}$, where $\Delta E$ is the energy difference between P and AP configurations corresponding to
the Zeeman energy splitting, $M_{\rm tot}$ is the total magnetic moment of the junction in the P
configuration, $g$ = 2 for graphene and $\mu_{\rm B}$=0.058 meV/T. A junction with two Fe-ZGNR leads of
length 10 nm and a Fe-AGNR spacer 2.4 nm in length and 7.2 {\AA} in width produces a $\Delta E$ of about 26
meV, which is comparable to room temperature $kT$. The junction with ideal edge termination of Fe atoms has a
$M_{\rm tot}$ of 415 $\mu_B$, thus resulting in a critical switching field as low as 0.54 T, which is much
less than that of the junction proposed by Mu\~{n}oz-Rojas \emph{et al.} \cite{Munoz-Rojas:2009}. In
experiments the possible formation of Fe wires of finite width at the Fe-ZGNR lead edges would enhance the
total magnetization and thus further reduce the effective switching field. This junction shows a big
advantage over the devices based on H-ZGNRs which mainly rely on ferromagnetic electrodes as well as local
magnetic control \cite{Kim:2008} or a huge switching magnetic field \cite{Munoz-Rojas:2009}. Current
nano-size lithographic techniques can provide the basis for fabricating and patterning of such junctions.

\section{\label{sec:5}Conclusion}

In summary, we have found that TM and NM terminated ZGNRs have completely different structures and magnetic
properties. The TM terminations transform semiconducting ZGNRs to metallic, while NM termination does not
affect the semiconducting nature of the ZGNR. ZGNRs with TM terminations and ferromagnetically coupled edges
states show a high spin polarization at the Fermi energy. We propose a magnetic junction with low, uniform
switching field using the Fe-ZGNRs as the ferromagnetic electrodes, and quantum transport calculations
indicate a large magnetoresistance is possible in such junctions. Our results outline a roadmap to observe
the magnetic order in graphene nanoribbon and hopefully stimulate experimental activities in the near future.

\emph{Note added in proof.} In an extended study we have examined the ribbon width dependence of magnetic
coupling between two edges of metal-terminated ZGNRs and AGNRs using projector augmented wave (PAW)
potentials within DFT, and the results will be published elsewhere.

\begin{acknowledgments}
This work was supported by US/DOE/BES/DE-FG02-02ER45995. The authors acknowledge DOE/NERSC and the UF-HPC
center for providing computational resources. C. Cao acknowledges support from NSFC (Grant No. 10904127).
\end{acknowledgments}


\begin{thebibliography}{99}
\bibitem{AGNR-junction} The Fe-AGNR in the FM state also shows a large spin polarization at the Fermi energy, as confirmed by our calculation.
\end{thebibliography}
\end{document}